\documentclass[twocolumn,a4paper]{article}
\usepackage{times,epsf,amsmath,amsfonts,amssymb}
\begin{document}
\noindent{\bf\Large Dynamics and flow-induced phase separation
  in polymeric fluids}\\\\
\noindent{\Large Peter ~D.  Olmsted}\\\\
The past few years have seen many advances in our understanding of the
dynamics of polymeric fluids. These include improvements on the
successful reptation theory; an emerging molecular theory of
semiflexible chain dynamics; and an understanding of how to calculate
and classify ``phase diagrams'' for flow-induced transitions.
Experimentalists have begun mapping out the phase behavior of wormlike
micelles, a ``living'' polymeric system, in flow: these systems
undergo transitions into shear-thinning or shear-thickening phases,
whose variety is remarkably rich and poorly understood.  Polymeric
ideas must be extended to include the delicate charge and composition
effects which conspire to stabilize the micelles and are strongly
influenced by flow.
\\\\
\noindent{\bf Address}\\
Department of Physics and Astronomy \& IRC in Polymer Science and
Technology, University of Leeds, Leeds LS2 9JT, UK; {\tt
  p.d.olmsted@leeds.ac.uk}
\section{Introduction} 
The field of non-linear rheology is roughly fifty years old and far
from mature. Non-linear fluids often display elastic effects, and have
effective viscosities which depend on stress or strain rate.
Polymeric liquids typically shear-thin \cite{doiedwards}, although
branched polymers thicken dramatically in extensional flow; colloidal
suspensions of platelike particles (clays) typically shear-thicken; and
solutions of surfactant (soap) molecules can shear-thicken \emph{or}
shear-thin.

Non-linear rheology and polymer dynamics are immense fields, and in
this short (subjective) review I will focus on a few subfields. I will
discuss recent advances in modelling the dynamics of flexible and
semiflexible polymer melts, including linear and complex topologies;
and then review progress in our knowledge of the surfactant
\emph{wormlike micelle} system, to which concepts from polymer
dynamics have been successfully applied. Unlike conventional polymers,
the micellar microstructure can change qualitatively in flow
conditions; these \emph{transitions} have many features in common with
equilibrium phase transitions, and have excited great interest.  A
good collection of results from a wide range of complex fluids may
found in \cite{challenges96}.
\section{Polymer Melts and Solutions}
The most accepted molecular model for the dynamics of flexible
entangled polymers has been the Doi-Edwards (DE) theory, based on
de~Gennes' reptation concept, in which polymers are envisaged to
occupy ``tubes'' that model entanglement constraints.  This has done a
reasonable job in predicting linear rheology, with a few notable
exceptions such as the failure to predict the scaling of the zero
frequency viscosity as $M^{3.4}$, although recent work suggests that
contour length \cite{MilnMcle98} fluctuations are a key to this
puzzle.  

Recent work in the linear regime includes applying the tube picture to
molecules with complex branched topologies \cite{tomreview}.  Star
polymers afford stringent tests of tube model ideas, because diffusion
is dominated by arm retraction in its tube, which is exponential in
the retraction potential.  Inclusion of higher order Rouse modes along
with ``dynamic dilution'' of the tube has led to remarkably good
agreement with experiment \cite{MilnMcle97}. In addition to star
polymers, molecular models have recently been developed for
progressively more complex topologies, paving the way for
understanding the flow behavior of industrially-important
long-chain-branched polymers, which strain-harden in extensional flow
while softening in shear flow \cite{tomreview}.

Although the DE tube picture works well in the linear regime, it has
several defects at high strain rates, particularly in steady shear.
For example: experiments show a slightly increasing plateau shear and
an increasing normal stress for strain rates ${\gamma}$ above an
inverse reptation time $\tau_r^{-1}$, while theory predicts a
\emph{decreasing} stress $\sigma\sim{\gamma}^{-1}$ and a constant
normal stress; and the DE theory predicts a high strain rate viscosity
which decreases with molecular weight, while experiments merge onto a
molecular weight-independent curve. The defect in DE theory is that,
as the tube representing entanglement constraints rotates into the
flow, the entrapped polymer feels a reduced stress and, at strain
rates above the inverse tube relaxation time $\tau_r^{-1}$, remains
oriented and presents a decreasing stress with increasing strain rate.
Although corrections due to tube stretching \cite{Marr96b} have
accounted for some problems in startup flows, this only applies near
${\gamma}\sim\tau_r^{-1}$ and still predicts a pronounced stress
maximum.  Another mechanism is needed to relax the chain and hence
increase the stress by providing more misaligned material for the flow
to ``grip''. The key is believed to lie in ``convected constraint
release'', whereby the entanglement (tube) mesh convects away at high
strain rates, leaving a relaxed coil. Early applications of this idea
\cite{IannMarr96,mead98} have cleared up several problems with the DE
theory, although the theory still predicts a slightly decreasing
stress with strain rate, so it should be regarded at provisional.

While moderately successful molecular theories of flexible polymer and
rigid rod dynamics have existed since the 70's, the study of
semiflexible polymers (in which $d<L_p<L$, where $d$ is the polymer
diameter, $L_p$ the persistence length, and $L$ the length) is quite
young, mainly due to the severe mathematical difficulties in treating
the bend degrees of freedom and length constraint. However, with
increasing attention being paid to biological polymers such as actin
\cite{MackJanm97}, a deeper understanding of the dynamics of
semiflexible polymer solutions is emerging. Direct imaging of tagged
fluorescent polymers is possible \cite{KSS94b}, and several techniques
have been developed for measuring elastic moduli, including direct
(torsional oscillator \cite{Janm+94}) and indirect (from various
optical techniques \cite{gittes97,mason97}). Experiments
indicate the vestiges of a plateau modulus, less pronounced than that
of flexible polymer solutions. For polymers with $L_p< L_e$, the
distance between entanglements or confinement constraints (or
deflection length \cite{odijk83}), one expects the behavior of
flexible solutions.  However, for $L_p>L_e$ one expects qualitatively
different effects due to the perturbation of bending modes by tube
constraints. While Odijk and Semenov \cite{odijk83,Seme86b} have
studied the dynamics and statistics of individual filaments, only
recently have molecular theories for the stress response of entangled
solutions emerged. Unlike flexible polymers, semiflexible chains have
a bending energy which maintains $L_p$, and one can distinguish
between longitudinal and transverse conformational changes. Two
pictures have emerged for the origin of elastic stress in concentrated
solutions: Isambert and Maggs \cite{IsamMagg96} argued that
semiflexible chains can slide along their tubes longitudinally, and
relaxation only occurs when transverse motions allow escape from the
tubes. MacKintosh \emph{et al.}  \cite{mkj95} argued that, if
longitudinal motion is suppressed, then the modulus is due to the
applied tension and the relaxation of bending modes (which are present
in the quiescent state due to thermal fluctuations). In the case of
solutions the former mechanism is expected to hold at times longer
than that on which chain tension can relax \cite{Magg97}.  These
pictures have been made more quantitative by Morse
\cite{morse98a}, who has developed a molecular theory at the
level of the Doi-Edwards theory and included the bending curvature
explicitly in the expression for the microscopic stress tensor.
 
\section{Flow instabilities in Wormlike Micelles}
DE theory predicts a bulk flow instability in polymer melts which has
not been seen; however, a suggestive instability known as the ``spurt
effect'' has been seen in extrusion, in which the throughput increases
dramatically above a critical pressure gradient, often accompanied by
a spatial pattern in the extrudate \cite{bagley58,vinogradov72}.
Current opinion is that this is a surface instability, although the
picture is not settled \cite{denn90}.  However, there is a polymeric
system which displays a well-documented bulk instability and has been
the subject of intense investigation in the past decade.

Certain aqueous surfactant solutions (\emph{e.g.} cetylpyridinium
chloride/sodium salicylate [CPCl/\-NASal]; cetyltrimethylammonium
bromide(CTAB)/\-NaSal) self-assemble into flexible cylindrical
micelles with an annealed length distribution that can encompass
polymeric dimensions (microns). These solutions comprise a surfactant
(\emph{e.g.} CPCl) and an ionizing salt (\emph{e.g.} NaSal) which
together determine micellar dimensions, flexibility, and interactions.
Salt and concentration effects are quite delicate, with Coulomb
interactions playing an important and poorly-understood role.  Micelle
reaction kinetics introduce additional timescales to the Rouse and
reptation times of conventional polymers: in the limit of fast
breaking times the stress relaxation of entangled micelles often obeys
a simple single exponential (``Maxwell fluid''), and properties can be
calculated quite confidently \cite{cates87,TurnCate92}, in good
agreement with experiment \cite{Khat+93}.  Some non-linear properties
can be calculated in this limit, and a maximum in the shear stress
(analogous the stress maximum in DE theory) is predicted at an inverse
relaxation time (the geometric mean of the reptation and breaking
times) \cite{cates90}, in quantitative agreement with experiment
\cite{rehage91}.

{\begin{figure}[!htb]
\epsfxsize=3.5truein 
\centerline{\epsfbox[50 40 660 475]{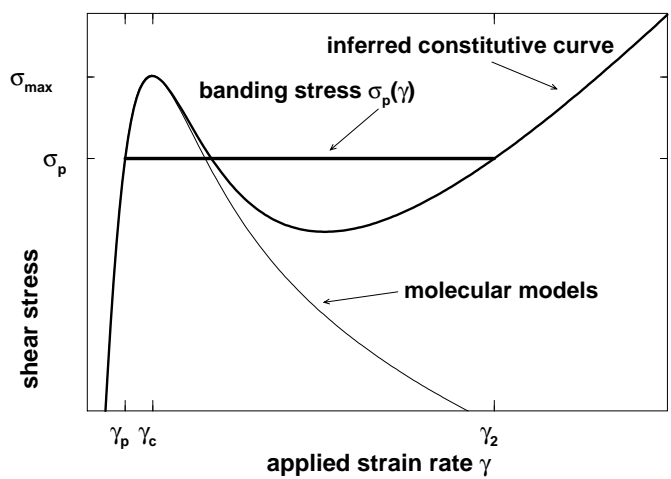}}
\caption{Model constitutive relation $\sigma(\gamma)$ for a
  shear-thinning complex fluid with a bulk instability (after
  Ref.~\cite{spenley93}). The thick curve shows the kind of curve
  envisioned for wormlike micellar solutions: this one is from the
  Johnson-Segalmann model \cite{malkus90}. Portions of the curve with
  negative slope are mechanically unstable. Both the Doi-Edwards
  theory for polymer melts and the Cates theory for micelles predict a
  constitutive curve qualitatively like the thin curve, with a
  continuously decreasing stress above a critical strain rate
  $\gamma_c$. Convected constraint release \cite{mead98} has reduced
  the instability in the polymer melt theory to a slight bump, while
  micellar theories still rely on an implicit solvent viscosity to
  stabilize the high strain rate branch (if there is indeed one).  The
  portion of the curve with negative slope is unstable. Experiments
  show that, for applied strain rates ${\gamma>}\gamma_p$, micellar
  solutions can phase separate under shear, attaining a unique steady
  state stress $\sigma_p({\gamma})$ in both phases, with a portion of
  the sample at a high shear rate $\gamma_2$ and a portion at the low
  shear rate $\gamma_1=\gamma_p$.  A dependence of $\sigma_p$ on
  ${\gamma}$ (not in this case) implies different compositions in the
  coexisting phases, so that \emph{different} constitutive curves are
  connected; and all properties change with mean strain rate
  \cite{schmitt95,olmstedlu97}.}
\label{fig:constit}
\end{figure}}
The non-linear rheology of micellar systems was first studied by
Rehage and Hoffman \cite{rehage91}, who discovered dramatic shear
thinning (analogous to the DE instability) and shear-thickening,
depending on the salt/surfactant/water composition.  The
shear-thinning systems were the first to be systematically studied
(see Figure~1).  Above a critical strain rate ${\gamma}_p$ an apparent
phase separation into macroscopic coexisting regions occurs, at a
reproducible and history-independent stress $\sigma_p$, in which the
high strain rate material is well aligned (typically birefringent) and
the low strain rate material remains relatively disordered.  The
underlying flow curve has the stress maximum $\sigma_{max}$ (predicted
by Cates \cite{cates90} for semi-dilute systems), while the composite
steady state flow curve has a plateau beginning at
$\sigma_p<\sigma_{max}$.  [It is important to note that micellar
systems have slow dynamics, and one can trap metastable states for
$\sigma>\sigma_p$ \cite{spenley93}.] This occurs in semi-dilute
systems, of order a few percent surfactant \cite{Makh+95}, or in more
concentrated systems (of order $30\%$) with a nearby equilibrium
nematic transition \cite{schmitt94,CCD95}.  In the former case the
dynamic instability is believed to be polymeric in nature
\cite{cates90}, while the latter may be due to nematic effects
(probably both effects are present).  No theories exist for nematic
transitions under shear in micelles, although recent work includes
phase diagrams for model rigid-rod suspensions in shear flow
\cite{olmstedlu97}, for which only a few results exist
[Ref.~\cite{mather97} reported a shear-induced nematic transition in a
liquid crystal polymer melt].

Shear banding can be inferred from rheological measurements and
directly observed optically.  Quantitative measurements include the
fraction of material and degree of alignment in the two phases,
inferred from neutron scattering \cite{BRL98}; and the velocity
profile, measured directly using magnetic resonance imaging
\cite{MairCall97}.  Shear-banding can incorporate different
concentrations in the two phases, which is expected when flow modifies
intermicellar interactions (as near a nematic transition) rather than
simply the micellar conformation (as might be expected in more dilute
systems). A signature of this is a slope in the ``plateau'' stress
with increasing mean strain rate \cite{schmitt95,olmstedlu97}, indeed
seen in concentrated solutions which often have an underlying nematic
transition \cite{Decr+95,BPD97,Capp+97,BRL98}.

Groups have begun investigating metastability. Berret \emph{et al.}
\cite{berret94b,Berr97} examined slow transients in 10-20\% CPCl/NaSal
solutions.  After increasing the strain rate into the two-phase region
the stress decayed slowly in time from the underlying constitutive
curve onto the stress plateau, with behavior $\sigma \sim
\exp-\left\{t/\tau\left({\gamma}\right)\right\}^{\alpha}, (\alpha=2)$,
which they interpreted as one-dimensional nucleation and growth.
Grand \emph{et al.} \cite{grand97} studied transients in more dilute
($\sim 1\%$) CPCl/NaSal solutions and found similar stress decays,
with $\alpha\simeq (2, 2.5, 3)$, and $\tau({\gamma})$ diverging above
or below (depending on composition) the strain rate ${\gamma}_p$ at
the onset of banding.  They also performed controlled stress
experiments, and discovered a stress $\sigma_{jump}>\sigma_p$, below
which the system remained on the low strain rate branch indefinitely
and above which the system eventually jumped to the high strain rate
branch.  Their data suggested that some compositions behave
``spinodal-like'' and others behave ``nucleation-like'' (as Berret's
did), but it is to early to completely embrace the language of
first-order transitions (given, \emph{e.g.}, $\sigma_{jump}$, which
has no equilibrium analog).

Fischer and Rehage showed how shear-thinning systems can be tuned by
changing surfactant and salt composition, from a shear banding
material to a material with a stress plateau (the rheological
signature of banding) but \emph{without} banding \cite{FiscReha97b}.
The shear and normal stresses apparently follow the \emph{Giesekus
  model}, which is one of the simplest non-linear constitutive
equations (comprising a Maxwell model with the simplest
stress-dependent relaxation time).  A molecular understanding for this
behavior is lacking.

{\begin{figure}[!htb]
\epsfxsize=3.5truein 
\centerline{\epsfbox[50 69 700 500]{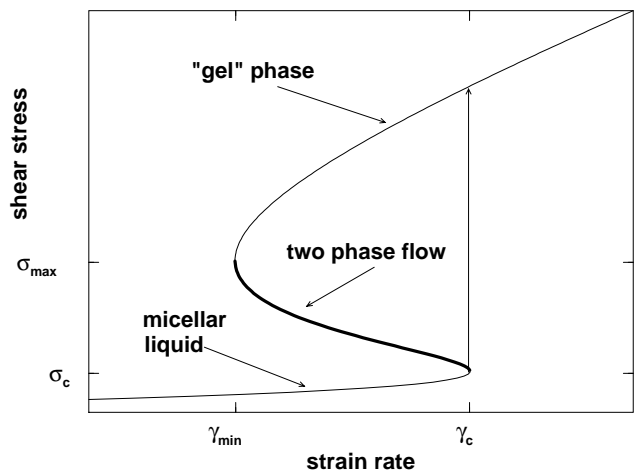}}
\caption{Qualitative constitutive curves for a shear-thickening
  system, such as the micellar system reported by Boltenhagen \emph{et
    al.} \cite{boltenhagen97b,HBP98}. The thin curves denote single
  phase relations (micellar liquid or a gel-like material); the thick
  curve denotes the composite flow curve along which the material
  phase separates for controlled stress $\sigma>\sigma_c$, and
  complete conversion to gel occurs for $\sigma=\sigma_{max}$. Upon
  controlling the strain rate the system traverses between the two
  branches at $\gamma_c$.  Note that, if compositions of the
  coexisting phases were identical, the controlled-stress composite
  flow curve would be vertical.}
\label{fig:constitthick} 
\end{figure}}
With critical micelle concentrations of order a few parts per million,
micelles entangle at astonishingly low dilutions.  Amazingly, systems
which shear-thin at concentrations of a few percent can undergo a
shear-\emph{thickening} transition at fractions of $<0.1\%$
\cite{rehage91,WFF98,HBP98}. This this shear-induced structure
(SIS) is still undetermined; early suggestions for the mechanism
included runaway micellar growth due to flow-alignment
\cite{catesturner92,BGB92}, but the observed strain rate (of order
inverse milliseconds) is much slower than the necessary micellar
reorientation time, of order $\mu\hbox{s}$.  It is probable that
charge, which controls the dramatic increase in micellar length for
concentrations near the overlap concentration \cite{MSP90}, plays an
important role.  Like the shear-thinning systems, macroscopic ``phase
separation'' occurs; Hu \emph{et al.} \cite{HBP98} found a gel-like
phase that forms upon increasing the applied stress, with the mean
strain rate decreasing (see Figure 2) as more material turns into gel,
and increasing again after complete conversion. The gel is observed to
fracture in flow, and slightly shear-thins.  Applying a strain rate
above the critical strain rate induces immediate complete conversion.
Attempts to visualize the SIS using cryo-TEM have given few clues to
the microstructure \cite{Kell+98}.  Note that coexistence in the
shear-thinning micelles occur under controlled strain rate conditions,
while coexistence in this thickening system occurs for controlled
stress; in both cases banding occurs in the radial direction,
indicating banding at a common shear stress. These differences may be
coincidences of the constitutive behaviors of the coexisting phases,
or due to whether stress or strain rate ultimately determines the SIS.

Berret \emph{et al.}  \cite{Berr+98} studied cethyltrimethylammonium
tosylate (CTAT) micelles, and found shear-thickening phase separation
under controlled \emph{strain rate} conditions above a critical strain rate
${\gamma}_c \sim\phi^{0.55}$ (an increase in ${\gamma}_c$ with $\phi$
was also found by Hu \emph{et al.}  \cite{HBP98}); this concentation
dependence remains unexplained. The composite curve $\sigma_p(\gamma)$
has a positive slope, in contrast to the \textbf{S} curve of
Ref.~\cite{HBP98}, possibly because the ``thick'' phase is not thick
enough; alternatively, phase separation along the vorticity direction
(at a common strain rate) would also be consistent with a positive
slope $d\sigma_p/d\gamma$ for the composite flow curve
\cite{olmstedlu97}.  The SIS is shear-thinning, displays an oriented
structure in neutron scattering, and does not have the extremely long
recovery times found in Ref~\cite{HBP98}. Qualitatively similar data
were reported for a CTAB-NaTOS solution by Harmann and Cressely
\cite{HartCres98}.  We finish our (incomplete) zoo of micellar
thickening transitions by mentioning yet another study on CPCl/NaSal:
revisiting early experiments by Rehage, Wheeler \emph{et al.}
\cite{WFF98} found spatio-temporal instabilities consisting of dark
and light oscillating vertical bands (in cylindrical Couette flow);
these accompany formation and destruction of new microstructure
(evident from turbidity), and are suggestive of a Taylor-Couette
elastic instability \cite{Lars92b}.

Athough wormlike micelles can have much simpler rheology than their
polymer cousins due to the frequent presence of a single relaxation
time in the fast breaking limit, this simplicity is delicate, and
strong flows can dramatically affect the micelle microstructure.  We
are far from a general molecular theory for these transitions, and do
not even \emph{know\/} the nature of the microstate in most cases.
Continuum constitutive models may provide some insight, although when
these models succeed we usually do not know why (\emph{e.g.} The
Giesekus model \cite{FiscReha97b}).  A popular constitutive model is
the local Johnson-Segalman model \cite{malkus90}, which is relatively
simple and displays the non-monotonic flow curve characteristic of
shear-thinning micelles. Numerical calculations
\cite{espanol96,greco97} resemble some startup experiments
\cite{Berr97}, and authors have attempted to determine plateau stress
$\sigma_p$ for the onset of banding in this model
\cite{malkus90,greco97}. It has become apparent that this, or any
local, model does not give a unique selected banding stress
\cite{spenley96}, and an additional assumption is necessary
\cite{porte97,olmstedlu97}.  There is growing consensus that non-local
(\emph{i.e.} gradient terms) contributions to constitutive equations
supply an unambiguous determination of the plateau stress
\cite{olmstedlu97}.
\section{Outlook}
Despite progress in understanding of flexible and semiflexible polymer
dynamics, there is no shortage of problems for the immediate future.
We lack a credible complete molecular understanding for \emph{any} of
these micellar flow-induced transtitions: such a model must presumably
include charge and concentration, as well as polymeric effects, to
account for the (still unknown, in most cases!) structural changes
under flow.  By comparing and contrasting living and non-living
polymers we may be able to extract important physics.  Besides these
systems, many other complex fluid systems undergo a variety of
flow-induced phase transitions, and it seems reasonable to hope that
this variety of transitions may be put on a common ground, akin to the
thermodynamics of equilibrium phase transitions.
\\\\
\noindent
I am indebted to J-F Berret, ME Cates, SL Keller, CYD Lu, FC
MacKintosh, TCB McLeish, DJ Pine, and G Porte for much discussion and
advice.


\end{document}